\documentstyle[epsf,12pt]{article}

\begin{document}
\bibliographystyle{unsrt}
\hfuzz=12pt

\font\twelvemb=cmmib10 scaled \magstep1
\font\tenmb=cmmib10
\font\ninemb=cmmib9
\font\sevenmb=cmmib7
\font\sixmb=cmmib6
\font\fivemb=cmmib5
\font\twelvesyb=cmbsy10 scaled \magstep1
\font\tensyb=cmbsy10
\font\sstwelve=cmss12
\font\ssnine=cmss9
\font\sseight=cmss8
\textfont9=\twelvemb
\scriptfont9=\tenmb
\scriptscriptfont9=\sevenmb
\textfont10=\twelvesyb
\def\bm{\fam9}
\def\bms{\fam10}

\newcommand{\DS}{\displaystyle}
\newcommand{\TS}{\textstyle}
\newcommand{\SS}{\scriptstyle}
\newcommand{\SSS}{\scriptscriptstyle}

\newcommand\zZtwelve{\hbox{\sstwelve Z\hskip -4.5pt Z}}
\newcommand\zZnine{\hbox{\ssnine Z\hskip -3.9pt Z}}
\newcommand\zZeight{\hbox{\sseight Z\hskip -3.7pt Z}}
\newcommand\zZ{\mathchoice{\zZten}{\zZten}{\zZeight}{\zZeight}}
\newcommand\ZZ{\mathchoice{\zZtwelve}{\zZtwelve}{\zZnine}{\zZeight}}

\mathchardef\sigma="711B
\mathchardef\tau="711C
\mathchardef\omega="7121
\mathchardef\nabla="7272

\newcommand{\e}{\epsilon}
\newcommand{\eps}{\epsilon}
\newcommand{\ee}{\varepsilon}
\newcommand{\vp}{\varphi}
\newcommand{\vphi}{\varphi}
\newcommand{\cphi}{\Phi}
\newcommand{\om}{\omega}
\let\oldvrho=\varrho
\newcommand{\vrho}{{\raise 2pt\hbox{$\oldvrho$}}}
\let\oldchi=\chi
\renewcommand{\chi}{{\raise 2pt\hbox{$\oldchi$}}}
\let\oldxi=\xi
\renewcommand{\xi}{{\raise 2pt\hbox{$\oldxi$}}}
\let\oldzeta=\zeta
\renewcommand{\zeta}{{\raise 2pt\hbox{$\oldzeta$}}}

\newcommand{\la}[1]{\label{#1}}
\newcommand{\ur}[1]{(\ref{#1})}
\newcommand{\ra}[1]{(\ref{#1})}
\newcommand{\urs}[2]{(\ref{#1},~\ref{#2})}
\newcommand{\eq}[1]{eq.~(\ref{#1})}
\newcommand{\eqs}[2]{eqs.~(\ref{#1},~\ref{#2})}
\newcommand{\eqss}[3]{eqs.~(\ref{#1},~\ref{#2},~\ref{#3})}
\newcommand{\eqsss}[2]{eqs.~(\ref{#1}--\ref{#2})}
\newcommand{\Eq}[1]{Eq.~(\ref{#1})}
\newcommand{\Eqs}[2]{Eqs.~(\ref{#1},~\ref{#2})}
\newcommand{\Eqss}[3]{Eqs.~(\ref{#1},~\ref{#2},~\ref{#3})}
\newcommand{\Eqsss}[2]{Eqs.~(\ref{#1}--\ref{#2})}
\newcommand{\fig}[1]{Fig.~\ref{#1}}
\newcommand{\figs}[2]{Figs.~\ref{#1},\ref{#2}}
\newcommand{\figss}[3]{Figs.~\ref{#1},\ref{#2},\ref{#3}}
\newcommand{\beq}{\begin{equation}}
\newcommand{\eeq}{\end{equation}}

\newcommand{\doublet}[3]{\:\left(\begin{array}{c} #1 \\#2
            \end{array} \right)_{#3}}
\newcommand{\vect}[2]{\:\left(\begin{array}{c} #1 \\#2
            \end{array} \right)}
\newcommand{\vectt}[3]{\:\left(\begin{array}{c} #1 \\#2 \\ #3
            \end{array} \right)}
\newcommand{\vectf}[4]{\:\left(\begin{array}{c} #1 \\#2 \\#3\\#4
            \end{array} \right)}
\newcommand{\matr}[4]{\left(\begin{array}{cc}
                   #1 &#2 \\
                   #3 &#4 \end{array} \right)}
\newcommand{\fracsm}[2]{{\textstyle\frac{#1}{#2}}}

\newcommand{\D}{{\cal D}}
\newcommand{\K}{{\cal K}}
\newcommand{\NC}{N_{\rm CS}}
\newcommand{\Ncs}{N_{\rm CS}}
\newcommand{\SU}{$SU(2)~$}
\newcommand{\Pmax}{P_{max}}
\newcommand{\tr}{\,{\rm tr}\,}
\newcommand{\Tr}{\,{\rm Tr}\,}
\newcommand{\ldef}{=}
\newcommand{\rdef}{=}
\newcommand{\simlt}{\stackrel{<}{{}_\sim}}
\newcommand{\simgt}{\stackrel{>}{{}_\sim}}

\newcommand{\nuH}{\nu_{\SSS H}}
\newcommand{\nuF}{\nu_{\SSS F}}
\newcommand{\nuf}{\nu_{\SSS f}}
\newcommand{\nut}{\nu_{\SSS t}}
\newcommand{\nuHold}{\nu_{{\SSS H}\,{\rm old}}}

\newcommand{\op}[1]{{\bf \hat{#1}}}
\newcommand{\opr}[1]{{\rm \hat{#1}}}
\newcommand{\bra}[1]{\langle#1\vert}
\newcommand{\ket}[1]{\vert#1\rangle}
\newcommand{\lsim}{\mathrel{\lower 2pt\hbox{$\stackrel{<}{\SS\sim}$}}}
\newcommand{\gsim}{\mathrel{\lower 2pt\hbox{$\stackrel{>}{\SS\sim}$}}}

\newcommand{\mucr}{\mu_{\rm \SSS crit}}
\newcommand{\Vpot}{V_{\rm pot}}
\newcommand{\Vpotm}{V_{\rm pot}^\mu}
\newcommand{\Tkin}{T_{\rm kin}}
 

 
\thispagestyle{empty}
\rightline{RUB-TPII-28/95}
\rightline{Januray 11, 1996}
\vspace{2cm}
\begin{center} {\Large\bf
Decay of high--density matter  \\ in the electroweak theory}

\vspace{3cm}
{\large\bf
J\"org Schaldach$^\diamond$, Peter Sieber$^\diamond,
\ $Dmitri Diakonov$^*$\footnote{
\noindent diakonov@lnpi.spb.su},\\
and Klaus Goeke$^\diamond$\footnote{
\noindent goeke@hadron.tp2.ruhr-uni-bochum.de}} \\
\vspace{20 pt}
\noindent
{\small\it 
$^\diamond$Inst. f\"ur Theor. Physik I\hskip -1.5pt I, 
Ruhr-Universit\"at Bochum, D-44780 Bochum, Germany\\
$^*$St.~Petersburg Nuclear Physics Institute, Gatchina, 
St.Petersburg 188350, Russia}
\end{center}
\vspace{1cm}
 
\abstract{High--density fermion matter is meta-stable due to the anomalous
non-conservation of baryon and lepton numbers in the electroweak theory.
The meta-stable state decays by penetrating the sphaleron barrier
separating topologically different vacua. The decay happens locally, and
results in an annihilation of twelve fermions, accompanied by a
production of gauge and Higgs bosons. We find numerically local bounce
solutions determining the decay rate, which are classical paths
in imaginary time, connecting two adjacent topological sectors.
We also follow the real-time evolution of the bosonic fields after the
tunneling and analyze the spectrum of the created bosons.}

\newpage

\noindent
{\bf 1.} 
It is well known that transitions between topologically different
sectors of the electroweak theory are accompanied by a change of
the fermion number \cite{tHooft}. 
The height of the separating sphaleron barrier is about
$E_{\rm sphal}\approx 10$ TeV \cite{AKY} and accordingly the tunneling rate
is extraordinarily small; it is exponentially suppressed by the factor
$\exp(-2 S_{\rm inst})\approx 10^{-153}$ with the instanton action
$S_{\rm inst}=8\pi^2/g^2\,,\ g\approx 0.67$. This prevents
baryon and lepton number violation from being an observable phenomenon
under ordinary conditions.
\par
This suppression is less
strong, or the transition is even unsuppressed at all, if the temperature
\cite{Kuzmin,Arnold,rub3}, the particle energy 
\cite{Ringwald,Espinosa,McLerran}, or the fermion density \cite{Rubakov,DiPet}
is large. Hence, fermion number violating processes might
have played a role in the earlier history of the Universe or could
perhaps be realized in future supercolliders. 
\par
In this paper we assume to have zero
(or very low) temperature but a macroscopic amount of fermions of a 
very high density
in thermal equilibrium. We describe them by the chemical potential $\mu$;
since we have zero temperature this is the energy up to which the fermionic
levels are occupied. In a sphaleron transition one level crosses the gap,
all others are shifted such that after the transition we have the same
spectrum again. But now one more level is occupied or depleted, depending
on whether the levels went up or down, hence the created or annihilated 
fermion has the energy $\mu$.
\par
The fermion number $N_{\rm f}$ of each doublet changes with the
Chern--Simons number $\Ncs$ of the bosonic fields as 
$\Delta N_{\rm f}=\Delta\Ncs$. It means that
the potential energy of the gauge and Higgs bosons is
shifted \cite{Rubakov,DiPet}:
\beq
\Vpotm=\Vpot+\mu\Ncs\ .
\la{vpotm}
\eeq
Some remarks are necessary here:
First, we have fixed the zero-point of the energy to the trivial vacuum 
with $\Ncs=0$.
Second, we may well neglect the change of the Fermi energy
as due to the creation or annihilation of a few fermions.
Third, $\Delta N_{\rm f}=\Delta\Ncs$ actually applies to all fermion
doublets, which also might have different chemical potentials. Therefore
in \eq{vpotm} we imply $\mu$ to be the sum of all these single
chemical potentials.
\par
The former degenerate ground states of the system with integer $\Ncs$ now have
energy $\mu\Ncs$ and hence are metastable (Fig.~1). They can decay by
tunneling through the sphaleron barrier to the adjacent vacuum
with lower energy. If $\mu$ exceeds a certain value $\mucr$,
the states become even unstable.
\par
\begin{figure} [ht]
\frenchspacing
\centerline{
\epsfxsize=6.in
\epsfbox[85 455 553 700]{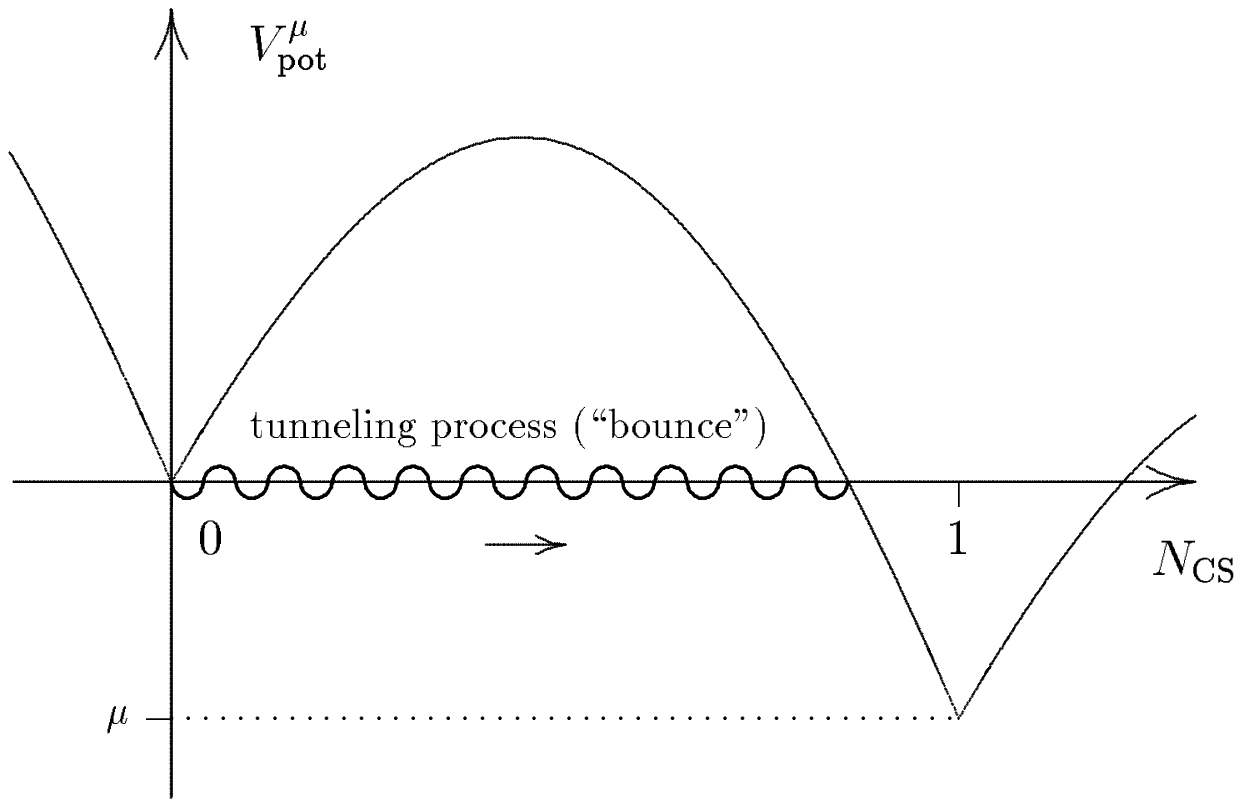}}
\nonfrenchspacing
\begin{quote}
\begin{quote}
{\it Fig.~1:\/} Schematic plot of the tunneling process between
two topological sectors.
\end{quote}
\end{quote}
\end{figure}

In \cite{DiPet}, arguments were given that the tunneling transition rate 
corresponds to the baryon number violation rate at high accelerator energies.
Thus the results of the present paper may also have
some relevance for the high energy transition rate.
\par
The decay of metastable ``false'' vacua was discussed by Coleman
already in 1977 \cite{Coleman}.
The transition rate per volume can be calculated in the semiclassical
WKB--approximation, it is of the form 
\beq
\Gamma/V=A\exp(-B)\;.
\la{rate}
\eeq
Here $B$ is twice the Euclidean action of the classical path
that connects the decaying state and the escape point at the other 
side of the barrier and minimizes this action. It is a movement 
in the inverted potential; the system leaves the initial state,
reaches the escape point with zero kinetic energy and would then move 
back the same way. Coleman called it ``the bounce'' . The prefactor
$A$ can be found from the small oscillation determinant about a bounce,
with one negative mode removed.
In the following we present the first numerical determination
of the bounce in electroweak theory.
 
\bigskip\noindent
{\bf 2.} 
We consider the minimal version of the standard electroweak theory
with one Higgs doublet in the limit of vanishing Weinberg angle.
In terms of dimensionless rescaled quantities the bosonic part of the
Lagrangian is
\beq
{\cal L}=m_W^4\Biggl[\frac{1}{g^2}\biggl(
-\fracsm{1}{4}F_{\mu\nu}^a F^{a\,\mu\nu}
+\fracsm{1}{2}(D_\mu \Phi)^\dagger(D^\mu \Phi)
- \fracsm{1}{32}\nu^2(\Phi^\dagger\Phi-4)^2\biggr)\Biggr]
\la{Lagr}
\eeq
with the covariant derivative
$D_\mu=\partial_\mu -i A_\mu\;,\ A_\mu=\frac{1}{2}A_\mu^a\tau^a$,
the field strength $F_{\mu\nu}=\frac{1}{2}F_{\mu\nu}^a\tau^a=
i[D_\mu,D_\nu]\;,\ F_{\mu\nu}^a=\partial_\mu A_\nu^a-\partial_\nu A_\mu^a
+\e^{abc}A_\mu^b A_\nu^c$, and the Higgs doublet
$\Phi={\cphi^+ \choose \cphi^0}$. $\nu=m_H/m_W$ is the ratio of Higgs
and gauge boson masses. 
The fermions are not considered explicitly here, but by
the modification \ur{vpotm} of the potential energy of the
bosonic fields. We work entirely in temporal gauge, $A_0=0$.
The potential and kinetic energy and the Chern--Simons index are
\begin{eqnarray}
\Vpot&=&\frac{m_W}{g^2}\int
d^3{\bf r}\,\left[\fracsm{1}{4}(F^a_{ij})^2+\fracsm{1}{2}(D_i\Phi)^\dagger
(D_i\Phi)+\fracsm{1}{32}\nu^2(\Phi^\dagger\Phi-4)^2\right]\;, \nonumber\\
\Tkin&=&\frac{m_W}{g^2}\int
d^3{\bf r}\,\left[\fracsm{1}{2} (\dot A_i^a)^2+\fracsm{1}{2}\dot\Phi^\dagger
\dot\Phi\right]\;, \\
\Ncs&=&\frac{1}{16\pi^2}\int d^3{\bf r}\,\left[
\epsilon_{ijk}\left(A_i^a \partial_j A_k^a
+ \fracsm{1}{3} \epsilon_{abc} A_i^a A_j^b A_k^c\right)\right]\ . \nonumber
\la{Eclass}
\end{eqnarray}
$\Ncs$ is only well-defined if the configuration space can be
identified with the sphere $S_3$, which requires the fields to be 
continuous at infinity. We fix them to the trivial vacuum there 
($A_\mu=0,\ \Phi={0 \choose 2}$).
In general, we will look for fields that minimize the action and expect them
to possess higher symmetries than completely arbitrary fields.
Therefore we assume the fields to have the spherical symmetry of the sphaleron:
\begin{eqnarray}
A^a_i(t,{\bf r})&=&\epsilon_{aij}
n_j\,\frac{1-A(t,r)}{r}+(\delta_{ai}-n_an_i)\,
\frac{B(t,r)}{r}+n_an_i\,D(t,r)\;,\nonumber\\
\Phi(t,{\bf r})&=&2\Bigl[H(t,r)+i G(t,r)\,{\bf n}\cdot{\bm\tau}\Bigr]
{\textstyle{0 \choose 1}}\;.
\la{hedge}
\end{eqnarray}
We assume the fields
to be continuous and differentiable everywhere and to yield
finite potential and kinetic energy. This poses some additional
conditions on the radial functions at $r=0$.
\par
For negative chemical potential $\mu$, the false vacuum
with $\Ncs=0$ can decay to the adjacent topological sector with
$\Ncs=1$, where the ground state has energy $\mu<0$. The minimal energy
barrier between the two sectors is given by the configurations of Akiba, 
Kikuchi, and Yanagida (AKY) \cite{AKY}
which are static fields that minimize the potential energy $\Vpotm$ for
given values of $\Ncs$ (Fig.~1).
\par
The barrier vanishes completely if $\mu$ exceeds a critical value $\mucr$.
{}From the condition that for $\mu>\mucr$ the potential $\Vpotm$ gets a
negative mode around the vacuum, one finds \cite{Rubakov,DiPet}
\beq
\mucr=\frac{16\pi^2}{g^2}\,m_W\;.
\la{mucrit}
\eeq
It is difficult to determine the bounce by solving numerically
the equations of motion as an initial
value problem. If we tried to find the propagation in time with a
Runge--Kutta method or something similar,
the system would always fall into some abyss of the potential $-\Vpotm$. 
Instead, we rather prefer to find a stationary point of the Euclidean action
\beq
S_E=\int_{-\infty}^{t_0}dt\left(\Tkin+\Vpotm\right)\;
\la{SE}
\eeq
directly, without using the equations of motion.
$t_0$ is the time when the system reaches the escape point;
since the lower bound is $-\infty$, we can choose an arbitrary value for $t_0$,
just as we fixed the origin for our spherical ansatz \ur{hedge} somewhere.
The corresponding translational invariance of the action does not influence
the factor $B$ in \eq{rate}, but contributes a space-time volume which
allows to have a transition probability per time and volume.
\par
$S_E$ is a functional of the five functions
$A,B,D,H,G$ of \eq{hedge}, depending on radial distance $r$ and time $t$.
Since the bounce has infinite extension both in space and in time,
we introduce new variables $x$ and $u$ which cover only finite intervals.
In practice, we use for example
\beq
r(x)=\lambda_r\arctan\left(\fracsm{\pi}{2}\,x\right)
\qquad{\rm and}\qquad
t(u)=\lambda_t\arctan\left(\fracsm{\pi}{2}\,u\right)\ .
\la{subst}
\eeq
Using ansatz \ur{hedge} and the substitution \ur{subst} we get
\begin{eqnarray}
S_E&=&\frac{S_{\rm inst}}{2\pi}\int_{-1}^{u_0}du\int_0^1 dx
\,\frac{1}{\om\vp}\biggl[\om^2\Bigl(\dot A^2+\dot B^2
+2r^2(\dot H^2+\dot G^2+\fracsm{1}{4}\dot D^2)\Bigr)
\la{SEhedge} \\
&+&(\vp A'+BD)^2+(\vp B'-AD)^2+2r^2(\vp H'+\fracsm{1}{2}GD)^2
\nonumber\\
&+&2r^2(\vp G'-\fracsm{1}{2}HD)^2
+\frac{1}{2r^2}(A^2+B^2-1)^2
+(H^2+G^2)(A^2+B^2+1) \nonumber \\
&+&2A(G^2-H^2)-4BGH+\fracsm{1}{2}\nu^2 r^2(H^2+G^2-1)^2 \nonumber \\
&+&2\rho\Bigl(D(A^2+B^2-1)+\vp BA'-\vp(A-1)B'\Bigr)\biggr]\nonumber
\end{eqnarray}
with
\beq
\rho=\frac{\mu}{\mucr}\;,\qquad\vp=\vp(x)=\left(\frac{dr}{dx}\right)^{-1}\;,
\qquad\om=\om(u)=\left(\frac{dt}{du}\right)^{-1}\;,
\eeq
and the dot and prime mean $\fracsm{d}{du}$ and $\fracsm{d}{dx}$, respectively.
In order to find a stationary point of $S_E$ numerically, we use a relaxation
method which was discussed by Adler and Piran in great detail \cite{Adler}.
The functional is put on a grid, let us say of size $n_u\times n_x$,
and the values of the five functions at the grid nodes are considered as
$5n_un_x$ independent variables.
Of course there is no unique way to discretize a functional, basically 
we followed the suggestions of \cite{Adler}. We postpone the details 
to a more comprehensive publication.
After putting some initial configuration on the grid, we sweep over it
by changing the single variables one after another. 
Each variable is changed according to the first iteration step 
of a Newtonian algorithm, which would find the stationary
point of $S_E$ if it was considered as a function of the variable in
question alone and all others were constants. This method is somewhat 
different from a procedure presented in \cite{Rubakov1}
where a similar problem in the context of techni-baryons in 
the Skyrme model is solved.
As mentioned above, the fields at $x=1\ (r=\infty)$ are fixed to
the trivial vacuum, and some conditions must be obeyed at the origin.
We take care that the initial configuration fulfills these properties
and that we do not loose them during the sweeps.
\par
In principle, one has to perform some hundred or thousand sweeps until the
situation is stable and the resulting configuration is a stationary point
of $S_E$.
But there are some difficulties which must be handled carefully to
get proper results:
\par
The main problem is that $\Vpotm$ is not bound, it can
be negative. Due to energy conservation the bounce itself cannot have
positive potential energy $-\Vpotm$ at any time,
its total energy $E_{\rm tot}=\Tkin-\Vpotm$ is constant and zero. 
But in its vicinity,
one can construct configurations which have positive potential $-\Vpotm$ for 
some time and which give a {\it lower} action than the bounce. So actually the
bounce is just a saddle point of $S_E$, it is a minimum only if we
restrict the potential energy $-\Vpotm$ to be non-positive.
So we enforce this restriction, simply by rejecting all changes of fields
rendering $-\Vpotm$ to be positive when we sweep. 
There are more sophisticated ways
to take into account invariances like energy conservation \cite{Kusenko},
but in our case
the simple remedy proved to be most effective.
Besides, we use energy conservation to check
if a given configuration is already close to the bounce
and if the numerical accuracy (e.g.~grid size) is sufficient.
\par
Another point is that the solution is not unique:
We want to describe the bounce between the topological sectors with
$\Ncs=0$ and 1; but this still allows so-called ``small''
time-independent gauge transformations which do not change the
Chern--Simons number.
We choose the $\Ncs=0$ vacuum to be the trivial one,
which fixes the gauge completely. Practically, we have a boundary
condition for the $u=-1$ edge of the grid, but due to the large
differences in the true time $t$ near this boundary,
the field variables are only weakly coupled there. We often observe
that the fields tend to another vacuum near the boundary, influenced
by the main configuration at larger $t$. In order to have a smooth
behavior, we sometimes apply a gauge transformation to the whole
grid except the close vicinity of the boundary which already
is in the trivial vacuum.
\par
Another degree of freedom is the choice of $t_0$, the time when the bounce
reaches the turning point, i.e.~the potential energy becomes zero again.
For the initial configuration this point lies on the right end of the grid,
during the sweeps we observe that it moves slowly to the left. 
For $t>t_0$ the potential
energy remains zero then (in the frame of our numerical accuracy).
Actually, this shift to the left never stops completely, but
finally the system moves as a whole, it just follows the zero-mode
due to the free choice of $t_0$.
\par
A more detailed description of the numerical technique will be given in 
a subsequent paper.
\par
In Table 1 we give our results for the bounce action and the 
Chern--Simons number
of the escape point for $\nu=1$ (i.e.~$m_H=m_W$) and several values of $\rho$.
The cases $\rho=0$  and 1 were not treated
numerically, but clearly for $\rho\to 1$ the action vanishes since
the barrier disappears. For $\rho=0$ it is known that we have the instanton 
action.
Actually, the finite vacuum expectation value of the Higgs field destroys
the scaling invariance of the instanton which we have in the case
of pure gauge fields, so the bounce is an instanton in the limit of 
size zero here \cite{Affleck1}. Only the introduction of the second 
massive parameter
$\mu$ allows the bounce to have finite size again in the other cases.
\vspace{0.7cm}
\[
\begin{array}{|c||c|c||c|c|c|c|c|}
\hline
\rho & S_E/S_{\rm inst} & N_{\rm CS}^{\rm esc} & N_W &
     N_H & E_W & E_H &  E_W+E_H \\
\hline
0.0 & 1.00  &  1.00   &         &         &         &         &          \\
\hline
0.2 & 0.82  &  0.85   &   22.9  &   1.9   &  0.370  &  0.020  &  0.391   \\
0.4 & 0.58  &  0.68   &   46.2  &   5.9   &  0.727  &  0.060  &  0.786   \\
0.6 & 0.34  &  0.51   &   67.1  &   9.6   &  1.093  &  0.102  &  1.195   \\
0.8 & 0.14  &  0.30   &  102.6  &  16.8   &  1.417  &  0.180  &  1.597   \\
\hline
1.0 & 0.00  &  0.00   &         &         &         &         &          \\
\hline
\end{array}
\]
\begin{quote}
{\it Tab.~1:\/} Bounce action, Chern--Simons number at the escape point,
and the number of gauge and Higgs bosons and their energies after the 
tunneling for several chemical potentials $\rho=\mu/\mu_{\rm crit}$; 
the energies are given in units of $\fracsm{8\pi^2}{g^2}\,m_W = 
\fracsm{\mu_{\rm crit}}{2}\,$. The mass of the Higgs boson is
fixed to $m_H=m_W\,$.
\end{quote}
 
One might expect that $S_E$ could have many different
stationary points, in other words, there could be several bounce
solutions with different escape points and values of the action.
As a matter of fact, for our numerics we used at least two distinct 
initial configurations for each value of $\rho$, but always found
the same solution.
\par
Fig.~2 shows the time dependence of the potential energy $-\Vpotm$ and
the Chern--Simons number $\Ncs$, with $t_0$ fixed to zero. In Fig.~3
we plotted $-\Vpotm$ versus $\Ncs$ and can compare this to the minimal
energy barrier. Obviously, the kinetic energy terms force
the system to take a path which deviates quite a bit from a possible
path through the AKY configurations minimizing the potential energy only. 
As one may expect, the deviation grows with the height of the barrier.
 
\begin{figure} [ht]
\frenchspacing
\centerline{
\epsfxsize=6.in
\epsfbox[85 313 553 700]{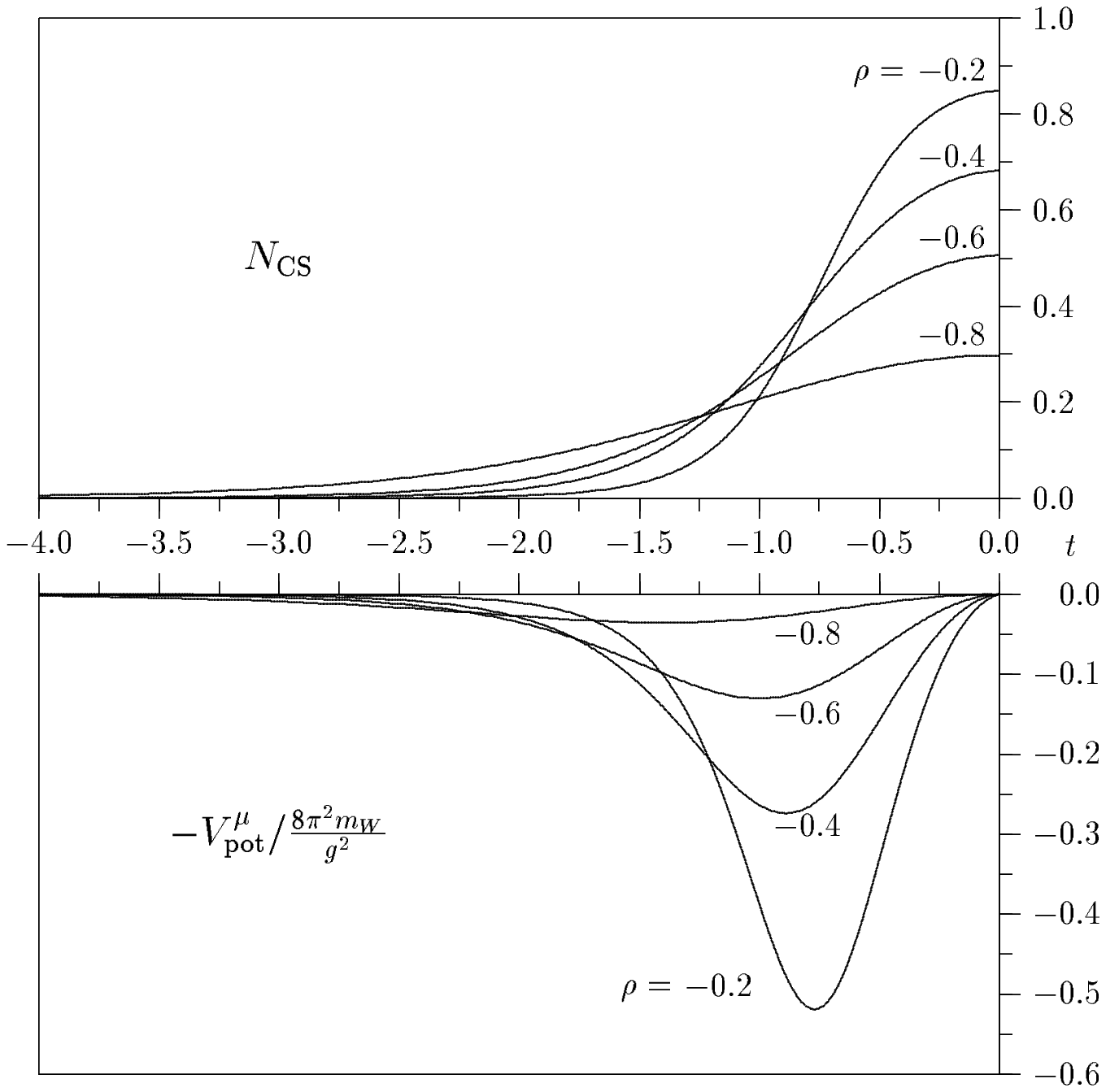}}
\nonfrenchspacing
\begin{quote} \begin{quote}
{\it Fig.~2:\/} Chern--Simons number $N_{\rm CS}$ and potential energy
$-V_{\rm pot}^\mu$ versus time $t$
(in units of $m_W^{-1}$) for different values of the chemical potential
$\rho=\mu/\mu_{\rm crit}\,$.
\end{quote} \end{quote}
\end{figure}
 
\begin{figure} [ht]
\frenchspacing
\centerline{
\epsfxsize=6.in
\epsfbox[85 425 553 700]{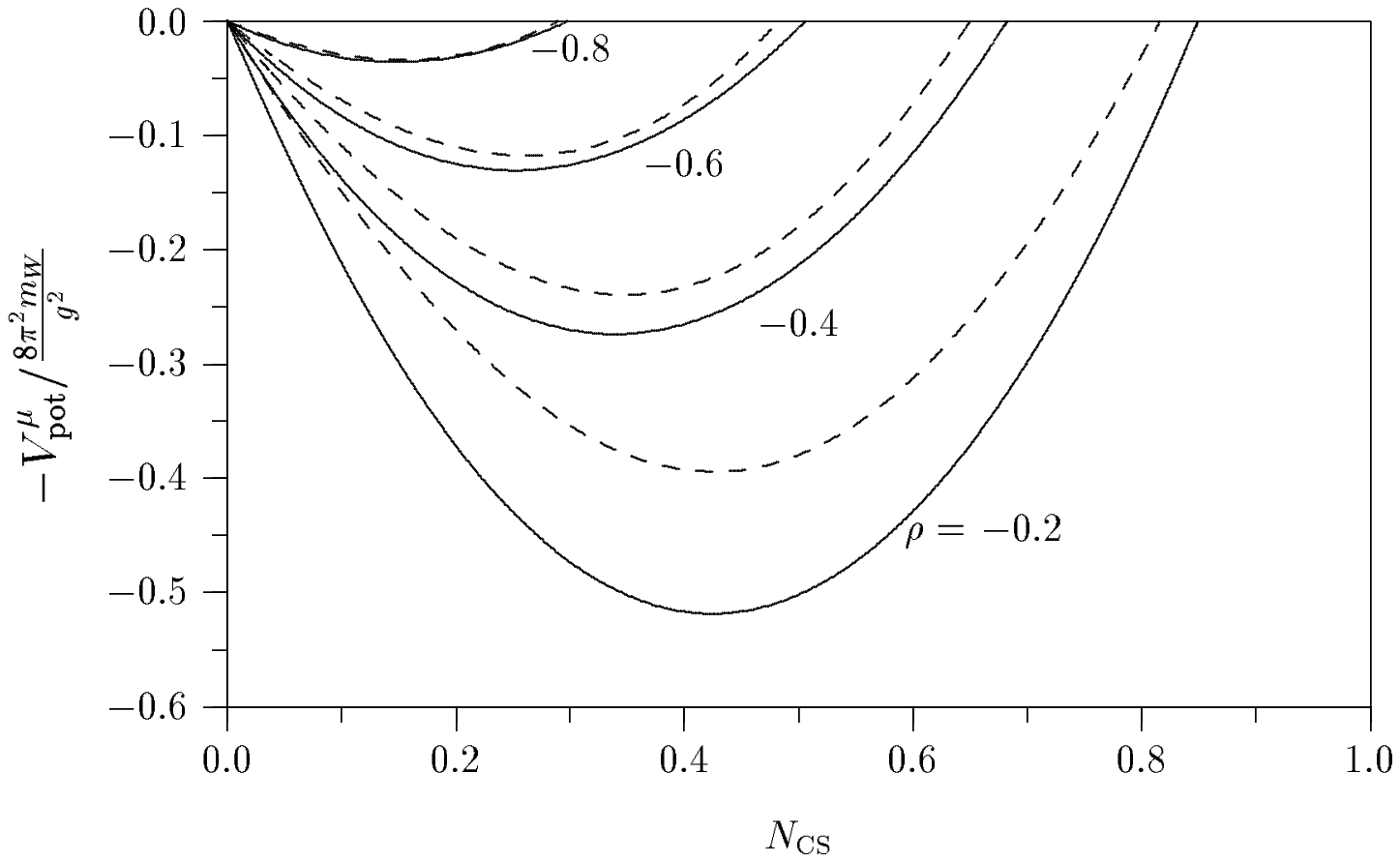}}
\nonfrenchspacing
\begin{quote} \begin{quote}
{\it Fig.~3:\/} The potential energy $-V_{\rm pot}^\mu$ versus the 
Chern--Simons number $N_{\rm CS}$ for various values of the chemical potential
$\rho=\mu/\mu_{\rm crit}$.
The solid lines correspond to the configurations which minimize the
action (\ref{SE},~\ref{SEhedge}) (bounce trajectories), the dashed 
lines are the minimal energy paths (AKY--configurations).
\end{quote} \end{quote}
\end{figure}
 
\bigskip\noindent
{\bf 3.} Next we want to examine the behavior of the system {\it after} 
the tunneling.
To this end, we take the (numerically found) escape point configuration and
let it propagate in the real Minkowskian time. We work on a two-dimensional 
grid again, but at this time we can solve the equations of motion using
a kind of Runge-Kutta method.
This was done in \cite{Hellmund,Cott} with a slightly disturbed sphaleron as 
initial configuration; we checked that our algorithm gives
the same results in this case.
\par
We have to distinguish two different kinds of behavior. The system can either 
stay in the topological sector where it came to after the tunneling or move 
classically over the barrier towards 
the next ground state with even lower energy.
For $|\rho|\simlt 0.2$, the system contains less energy than required to cross
the next barrier, so that it will definitely stay in its topological sector.
For values of $|\rho|$ between about 0.2 and 0.9 the system has enough total 
energy so that in priciple it could move over the barrier. The system, however,
stays in its sector, because the energy is dissipated along
the exitation modes in such a way that the system does not find a collective 
path towards the next vacuum. Finally for  $|\rho|\simgt 0.9$ the system
actually moves classically into the next sector, and subsequently also
crosses the following barriers, i.e.~it keeps moving into the direction
of growing $N_{\rm CS}$.
\par
In the case when the system stays in its topological sector,
we see that after some short time the potential and kinetic energies
become almost constant. For the transition described above, from $\Ncs=0$ to 1,
we observe very small oscillations of $\Tkin$ around $|\mu|/2$
and of $\Vpotm$ around $-|\mu|/2$, which is in accordance with the virial
theorem. The energy is concentrated near a sphere which expands with 
about 95\% of the speed of light.
$\Ncs$ becomes almost constant after a while, with small variations; 
but this constant is not the integer number which classifies the 
topological sector. Nevertheless, the fermion number is changed by one
for each doublet: The equation $\Delta N_{\rm f}=\Delta\Ncs$
is not valid here because in the Minkowskian setup
the non-vacuum fields cannot be restricted to a finite region
\cite{Farhi}.
\par
Finally, we find that the time-averaged fields 
($\fracsm{1}{b-a}\int_a^b\cdots\,dt$ with $b\to\infty$)
form a static vacuum configuration $\bar A^a_i,\bar\Phi$ with $\Vpot=0$ and
integer $\Ncs$, and that at larger times the fields can be considered as small
fluctuations around that vacuum. This allows to analyze the particle contents
of the bosonic fields, we can perform a spectral decomposition in the basis
of free particle modes \cite{Hellmund}. We apply a gauge transformation to
the fields which converts $\bar A^a_i,\bar\Phi$ to the trivial vacuum first,
hence we work with fluctuations $a,b,d,h,g$ around the latter one.
The basis is found by solving the linearized
equations of motion; basically we follow \cite{Hellmund} here,
but use $\Vpotm$ instead of $\Vpot$. The free particle modes
in our spherical ansatz are
\begin{eqnarray}
a_k(t,r)&=&\fracsm{3}{2}\Bigl(\beta_1^k(t)-\beta_2^k(t)\Bigr)\,r\,j_1(kr)
\la{basis} \\
b_k(t,r)&=&\Bigl(\beta_0^k(t)-\fracsm{1}{2}\beta_1^k(t)
-\fracsm{1}{2}\beta_2^k(t)\Bigr)\,r\,j_2(kr) \nonumber \\
&+&\Bigl(\beta_0^k(t)+\beta_1^k(t)+\beta_2^k(t)\Bigr)\,r\,j_0(kr) \nonumber \\
d_k(t,r)&=&\Bigl(-2\beta_0^k(t)+\beta_1^k(t)+\beta_2^k(t)\Bigr)\,j_2(kr)
\nonumber \\
&+&\Bigl(\beta_0^k(t)+\beta_1^k(t)+\beta_2^k(t)\Bigr)\,j_0(kr) \nonumber \\
h_k(t,r)&=&h^k(t)\,j_0(kr) \nonumber \\
g_k(t,r)&=&-\fracsm{3}{2}\,\beta_0^k(t)\,k\,j_1(kr) \nonumber
\end{eqnarray}
with
\beq
\beta_i^k(t)=\beta_i^k\sin(\omega_it+\alpha_i)\;,\qquad\qquad
h^k(t)=h^k\sin(\Omega t+\gamma)\;,
\la{sinus}
\eeq
\beq
\omega_0^2=1+k^2\;,\qquad\omega_{1,2}^2=1\pm2\rho k+k^2\;,\qquad
\Omega^2=\nu^2+k^2\;.\nonumber
\eeq
The $j_s$ are the spherical Bessel functions. For each time $t$ and
momentum $k$ we can get the functions $\beta_i^k(t)$ and $h^k(t)$
from radial ``Fourier'' integrals; a comparison with \ur{sinus}
yields the coefficients $\beta_i^k$ and $h^k$.
\par
In the linearized theory, the total energy is
\beq
E=E_W+E_H=m_W\int_0^\infty dk\;n_i(k)\,\omega_i(k)+
m_W\int_0^\infty dk\;n_H(k)\,\Omega(k)
\la{Elin}
\eeq
with the spectral densities
\begin{eqnarray}
n_0(k)&=&\frac{9\pi^2}{g^2k^2}\,(k^2+1)\,\omega_0\,(\beta_0^k)^2
\la{spdens} \\
n_1(k)&=&\frac{9\pi^2}{g^2k^2}\,\omega_1\,(\beta_1^k)^2 \nonumber \\
n_2(k)&=&\frac{9\pi^2}{g^2k^2}\,\omega_2\,(\beta_2^k)^2 \nonumber \\
n_H(k)&=&\frac{4\pi^2}{g^2k^2}\,\Omega\,(h^k)^2\ . \nonumber
\end{eqnarray}
Fig.~4 shows the spectral densities for the gauge and Higgs particles,
$n_W=n_0+n_1+n_2$ and $n_H$. The total particle numbers
\beq
N_W=\int_0^\infty dk\;n_W(k)\qquad{\rm and}\qquad
N_H=\int_0^\infty dk\;n_H(k)
\la{Ngh}
\eeq
and the energies $E_W$ and $E_H$ are listed in Table 1. The good
agreement of $E_W+E_H$ with the energy $\mu$ gained from the fermions
shows that the description in terms of the linearized theory is fully
justified.
 
\begin{figure} [ht]
\frenchspacing
\centerline{
\epsfxsize=6.in
\epsfbox[85 295 553 700]{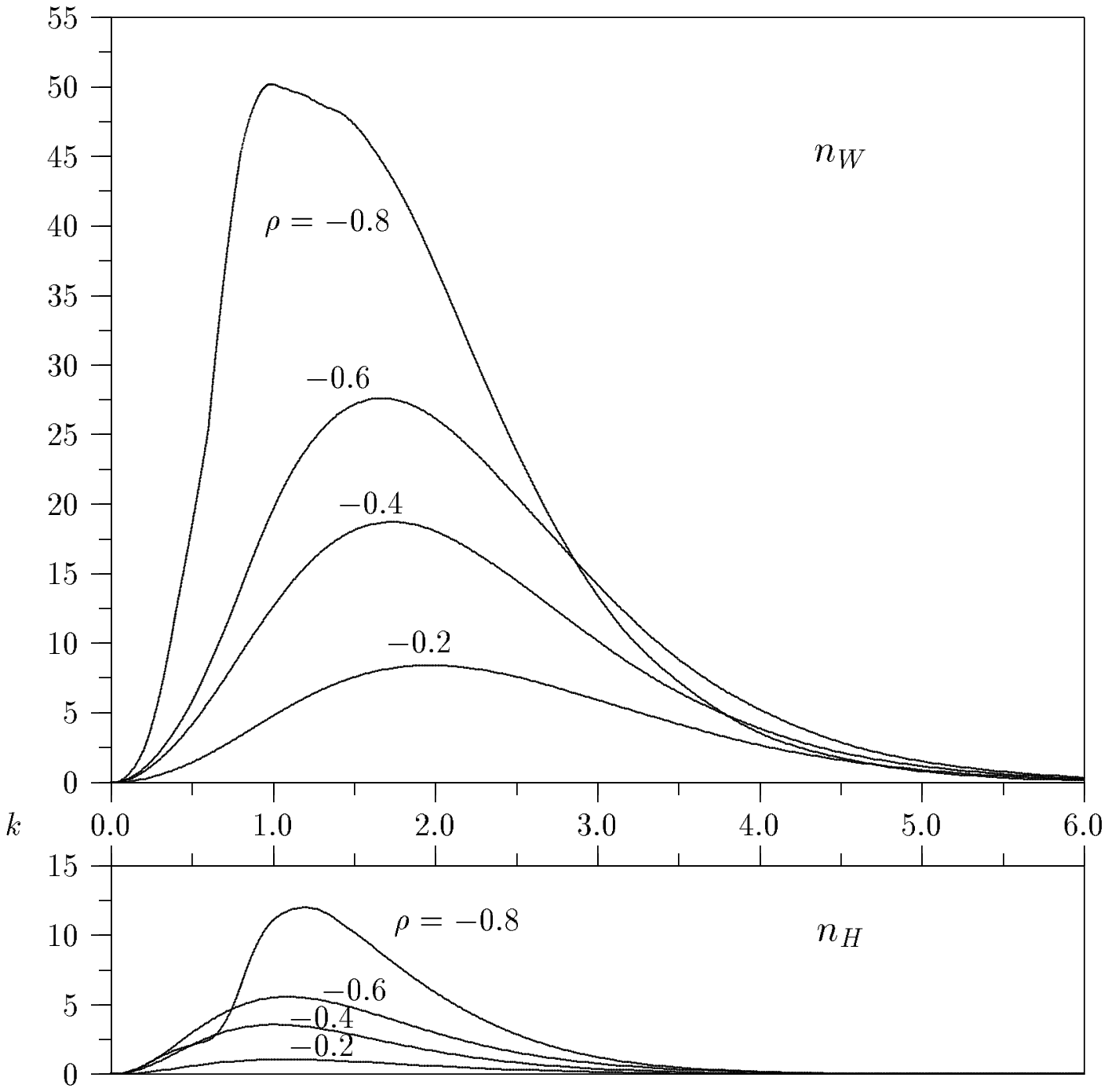}}
\nonfrenchspacing
\begin{quote} \begin{quote}
{\bf Fig.~4:} The spectral densities $n_W(k)$ and
$n_H(k)$ (in units of $m_W^{-1}$) for various values of the chemical
potential $\rho=\mu/\mu_{\rm crit}\,$.
\end{quote} \end{quote}
\end{figure}
 
\bigskip\noindent
{\bf 4.}
In summary, our work falls into two main parts, both of numerical character.
First, we found the classical bounce trajectory between two topologically
different sectors as a stationary point of the discretized Euclidean action
on a grid in time and the radial space-coordinate.
A careful treatment is required since this is a saddle point rather than
a minimum, and numerical problems due to the gauge freedom must be handled 
properly.
Second, we examine the
real-time behavior of the fields after the tunneling, which
gives some hints about the possible
bosonic signature of baryon number violating processes.
Our ansatz assumed a background of high fermionic density, but
the results should also be interesting for the case of
few particles with very high energy.
 
We are grateful to C.Weiss for pointing our attention to the use of
relaxation methods (\cite{Adler}), and thank P.Pobylitsa, M.Polyakov,
and V.Petrov
for numerous discussions. The work has been supported in part by the Deutsche
Forschungsgemeinschaft, the RFBR grant 95-07-03662,
and the joint Russian government--International Science Foundation
grant R2A300.

\end{document}